\documentclass[pra, twocolumn]{revtex4}
\usepackage{amstext,amsmath,amssymb,amsfonts}
\usepackage{bm,bbm,euscript}

\def\1{{{\mathbbm 1}}}
\def\6{\langle}
\def\9{\rangle}

\def\be{\begin{equation}}
\def\ee{\end{equation}}
\def\bay{\begin{array}}
\def\ear{\end{array}}

\def\arr{\rightarrow}

\def\tr{{\mathrm{tr}}\,}

\def\eL{\EuScript{L}}

\def\mout{{\mathrm{out}}}
\def\mrin{{\mathrm{in}}}

\begin{document}

\title{Entanglement, discord and the power of quantum computation}
\author{Aharon Brodutch}
\email{aharon.brodutch@mq.edu.au}
\author{Daniel R. Terno}
\email{daniel.terno@mq.edu.au}
\affiliation{Department of Physics \& Astronomy, Faculty of Science, Macquarie University, NSW 2109, Australia}

\begin{abstract}

We show that the ability to create entanglement is necessary for execution of bipartite quantum gates even when they are applied to unentangled  states and create no entanglement.  Starting with a simple example we demonstrate  that to execute such a gate bi-locally the local operations and classical communications (LOCC) should be supplemented by shared entanglement. Our results point to the changes in
quantum discord, which is a measure of quantumness of correlations even in the absence of
entanglement, as  the indicator of  failure of a LOCC implementation of the gates.

\end{abstract}
\maketitle

The question ``What makes a quantum computer tick?" goes back to the early discussions of quantum algorithms \cite{jos}. Two different explanations of  the  speed-up of  quantum algorithms are centered on the two fundamental aspects of quantum theory:  superposition of quantum states  and their entanglement  \cite{book,kendon}.

The latter view is supported by the make-up of a universal set of gates \cite{book}. To run a quantum computation it is sufficient  to execute certain one-qubit gates and one entangling gate, such as  a two-qubit controlled-{\it NOT} ({\it CNOT}). An entangling gate turns a generic non-entangled input into an entangled output. On the other hand, any pure-state quantum computation that utilizes only a restricted amount of entanglement can be efficiently simulated classically \cite{vidal}.

 According to the alternative view, it is  a  superposition of all possible computational paths in a quantum computer that is responsible for  a speed-up, while entanglement may be just incidental.  Indeed, the algorithm DQC1  demonstrates such a speed-up without entanglement \cite{powerof1,dqc1}.

We show that entanglement is required for the implementation of bipartite gates, even if they operate on a restricted set $\eL$ of unentangled input states that are transformed into unentangled outputs. This  remains true when the set is chosen to contain only mixtures of some pure states, and not their coherent superpositions.

 A distributed implementation of a gate is a natural setting to study the effects of entanglement.  A(lice) and B(ob) execute a bipartite gate $U$ using local operations and different shared resources. We show that under quite general assumptions $U$ can be implemented bi-locally  on $\eL$ only if Alice and Bob share some entanglement. The build-up of quantum correlations other than entanglement as $\rho^\mrin$ is transformed into $\rho^\mout$ indicates the demand for shared entanglement. The correlations are quantified by quantum discord \cite{disc-int}.

We first introduce quantum discord and review some of its properties, then present a simple example and follow with  general results.

\textit{Discord }  is defined through the difference in the generalizations of two expressions for the classical mutual information,
\be
I(A:B)=H(A)+H(B)-H(AB), \label{mi}
\ee
and
\be
J(A:B)=H(A)-H(A|B)=H(B)-H(B|A), \label{mj}
\ee
where $H(X)$ is the Shannon entropy of the probability distribution $X$, $H(Y|X)$ the conditional entropy of $Y$ given $X$, and $H(XY)$ is the entropy of  a joint probability distribution \cite{ct}. The two classical expressions are equivalent. The quantum measurement procedure $\Lambda$ on a state $\rho$ leads to a probability distribution $X_\rho^\Lambda$. The von Neumann entropy $S(\rho_X)=-\tr\rho_X\!\log\rho_X$ replaces the Shannon entropy \cite{peres}, but the  conditional  entropy now explicitly depends on the measurement procedure \cite{dat,disc-int} and the optimization goal it tries to achieve. For our purposes it is enough to assume that the measurement $\Pi_A$ on Alice's subsystem is represented by a complete set of orthogonal projections, and the optimization is chosen to lead to the discord measure $D_2$ \cite{z03,bt1}. Then
\be\label{J2}
J^{\Pi_A}_2(\rho):=S(\rho_B)-S(\rho_B|\Pi_A)+S(\rho_A)-S(\rho^{\Pi_A}_A),
\ee
where the averaged post-measurement state of $A$ is
\be
\rho^{\Pi_A}_A=\sum_a p_a\Pi^a_A,\qquad p_a=\tr\rho_A\Pi_A^a,
\ee
 (classically the last two terms in (\ref{J2}) cancel  out), and  the conditional entropy of the post-measurement state of $B$,
 \be
S(\rho_B|\Pi_A):=\sum_a p_a S(\rho_{B|\Pi^a_A}).
\ee
is the weighted average of the entropies of the states
\be
\rho_{B|\Pi_A^a}= \tr_A(\Pi_A^a\otimes\1_B \rho\Pi_A^a\otimes\1_B)/p_a
\ee
that correspond to the individual outcomes. Finally,
the discord is
\be
D_2^A(\rho):=\min_{\Pi_A}[H(A_\rho^\Pi)+S(\rho_B|\Pi_A)]-S(\rho_{AB})\geqslant0. \label{def_d2}
\ee
Discord has a number of interesting properties and applications \cite{z03,bt1,discp,vl10}.
We  use the  property \cite{bt1}:
\be
D_2^{\Pi_A}(\rho)=S(\rho^{\Pi_A})-S(\rho)\geqslant D_2^A(\rho), \label{ent_inc2}
\ee
which holds for any set $\Pi_A$ that induces the averaged post-measurement state $\rho^{\Pi_A}$.

\textit{Gate implementation}. We investigate a bi-local implementation of the gate  $U$ on a restricted set  $\eL$. Alice and Bob can perform arbitrary local operations  and measurements on their respective qubits,  are allowed to exchange unlimited classical messages, but have no shared entanglement.  While it is just a standard LOCC paradigm, we point out one important feature of the reduced dynamics of the system.

 The measurements are represented by arbitrary local positive operator-valued measures (POVM), so Alice's measurement is given by a family of positive operators of the form $E^\mu_A=\Lambda_A^\mu\otimes\1_B$, $\Lambda_A^\mu>0$, $\sum \Lambda_A^\mu=\1_A$. At each stage the operations and measurements are integrated together with the help of an ancilla, which can be further divided into two parts $A'$ and $A''$, as in \cite{tea99}. The measurement is accomplished in two stages: first some unitary operation $U_{AA'A''}$ is applied to the entire system, and then a projective measurement $\Pi^a$, $a=1,\ldots \dim A''$, $\Pi^a\Pi^b=\Pi_a\delta_{ab}$ is done on the system $A''$. Depending on the outcome, a unitary $U_{AA'}(a)$ is applied to the remaining part $AA'$. While the entire evolution of $A$ is completely positive, that is, $\rho_A^\mrin\mapsto\rho_{A|\Pi^a}\mapsto\rho_A^\mout=\sum_\mu K_\mu\rho_A^\mrin K_\mu^\dag$ for some set of Kraus matrices $K_\mu$ \cite{book}, the evolution of a post-measurement state $\rho_{A|\Pi^a}\mapsto\rho_A^\mout=\tr_{A'}U_{AA'}(a)\rho'_{AA'}U^\dag_{AA'}(a)$ generally depends on the correlations between $A$ and $A'$ and may be not completely positive \cite{dhk}.

 \textit{Example:  A CNOT gate}. This gate can be performed bi-locally by Alice and Bob if they share one ebit of entanglement per gate use \cite{eisert00}. In our example
  Alice and Bob share an unknown state from the known list $\eL$ and try to implement the CNOT gate by LOCC. It is obvious that if the set $\eL$ is  locally distinguishable, then the gate  can be  implemented by LOCC. It is also obvious that if the action  creates entanglement, the implementation fails. However,  absence of entanglement is not sufficient.

Consider the set $\eL$ in Table~I.
\begin{table}[htbp]
   \centering
    \caption{Four inputs and outputs for the CNOT gate}     \label{input4}
\begin{tabular}[v]{cl|cl}\hline\hline
\# & State & \# & {State} \\ \hline
$a$&$|1\9|Y_+\9 \arr i|1\9|Y_-\9$ & $c$ &$|Y_+\9|X_-\9 \arr |Y_-\9|X_-\9$\\
$b$&$|0\9|Y_+\9 \arr|0\9|Y_+\9$ & $d$ & $|Y_+\9|X_+\9 \arr |Y_+\9|X_+\9$\\\hline\hline
\end{tabular}
\end{table}

\noindent Here $\sigma_y|Y_\pm\9=\pm|Y_\pm\9$, $\sigma_x|X_\pm\9=\pm|X_\pm\9$, where $\sigma_{x,y,z}$ are Pauli matrices.

We  demonstrate that ability to implement the CNOT gate on $\eL$ without shared entanglement makes it possible to  unambiguously discriminate between these non-orthogonal states using just one input copy, which is impossible \cite{peres}. Without specifying the local operations of Alice and Bob we classify them  according to their action on the sate $|Y_+\9$. An operation  $\Phi$ is \textit{flipping} (F) if up to a phase $\Phi(|Y_+\9)=|Y_-\9$, \textit{non-flipping} (N) if  $\Phi(|Y_+\9)=|Y_+\9$, and  is undetermined otherwise.

Knowing the  operation type allows  Alice and Bob to narrow down the list of  possible inputs: For example, Bob's F is incompatible with having the input $b$, while for Alice's operation not to have a definite type excludes both $c$ and $d$. The list of possible inputs if both operations are of a definite type is presented in Table~II. If one of the performed operations  is neither F nor N, then the type of other operation allows to determine the input uniquely.

\begin{table}[htbp]
   \centering
    \caption{Possible inputs}
 \begin{tabular}[v]{c|cc} \hline\hline
Alice & \multicolumn{2}{c}{ Bob}  \\
 & F & N \\ \hline
 F & $\left\{\bay{cc} a & c \\   &  \ear\right\}$ & $\left\{\bay{cc}   & c \\ b &  \ear\right\}$ \\

 N & $\left\{\bay{cc} a &   \\   & d\ear\right\}$ & $\left\{\bay{cc}   &   \\ b & d\ear\right\}$ \\\hline\hline
 \end{tabular}
\end{table}
Any pair of outputs can be reset to their original input state by local unitaries and resent through the gate. For example, if the overall operation is of the FF type, the operation $\sigma_z^A\otimes\sigma_x^B$ will transform the outputs $\psi_a'=|1\9|Y_-\9$ and $\psi_c'=|Y_-\9|X_-\9$ into the inputs $\psi_a$ and $\psi_c$, respectively.

The operations that implement the gate on its second run may be  the same or different from the operation in the previous run. If the gate's  design allows a finite probability of having a different operation type, it will be realized after a finite number of trials. This other type (FN or NF in the preceding example) will uniquely specify  the input. If  a particular pair of inputs is always processed by the same type of operations, then the gate can be used to unambiguously distinguish between one state from this pair and at least one of the two remaining states in a single trial.  \hfill$\blacksquare$

\textit{Definition}. A  bi-local implementation $G$ of a gate $U$ on some (finite) set of unentangled states $\eL=\{\rho_i^\mrin\}_{i=1}^N$ (and their convex combinations) is a completely positive map that is implemented by local  operations on the subsystems $A$ and $B$, performed separately, that are assisted by unlimited classical communication such that for any state $\rho_i\in \eL$
\be
G(\rho_i^\mrin)=\sum_kK_k\rho_i^\mrin K_k^\dag\equiv U\rho_i U^\dag=\rho_i^\mout.
\ee

Successful implementation of the gate on pure  inputs guaranties that it is ``reversible", with the dual map \cite{karol} playing the role of the inverse.

 \textit{Property 1.} The dual map $G^+(\rho):=\sum_k K_k^\dag \rho K_k$ satisfies
\be
\rho_i^\mrin=G^+(\rho_i^\mout)
\ee
for all pure input states $\rho_\psi\in\eL$.

\noindent \textit{{Proof:}} Since $\rho_\psi^\mout=G(\rho_\psi^\mrin)=U\rho_\psi U^\dag$ is pure, using the Hilbert-Schmidt inner product we see that
\be
1=\6\rho_\psi^\mout,\rho_\psi^\mout\9=\6\rho_\psi^\mrin, G^+(U\rho_\psi^\mrin U^\dag)\9,
\ee
hence $G^+$ acts as an inverse for all allowed pure inputs and their convex combinations.  \hfill $\blacksquare$

It is straightforward to see that if we restrict local operations to projective measurements and unitaries, then the zero discord becomes a necessary criterion for such implementation's success. Namely, since entropies of initial and final states are the same, but a local measurement on a state of non-zero discord increases it according to Eq.~\eqref{ent_inc2}, we reach a contradiction.

A symmetrized version of the discord is used in what follows:
\be
D_2(\rho):=\min[D_2^A(\rho),D_2^B(\rho)]\neq 0.
\ee
Unlike the exact value of discord that can be calculated analytically only in special cases, it is straightforward to check weather the discord is zero or not \cite{bt1}. Moreover, sates of zero discord (say, $D_2^A=0$) are of the form
\be
\rho=\sum_a p_a\Pi_A^a\otimes\rho^a_B, \qquad  p_a\geq 0, \quad \sum_a p_a=1. \label{zerodisc}
\ee

Now we consider different bi-local implementation of $U$.  Assume first that the the set of possible inputs $\eL$ includes the maximally mixed state (i.e. the gate is unital, $G(\1)=\1$), and at least one pure state that we  write as $|00\9$ . Also restrict the allowed local operations to  completely positive (CP)  maps (this is realized, in particular,  if at each stage the ancilla is entirely consumed by the measurement, i.e., $\dim A'=0$).

\textit{Lemma 1.} If a set $\eL$ contains   one pure product state ($|00\9$) and the maximally mixed sate ($\1/4$) in $\eL$, and the action of $U$ is realized by  local operations restricted to arbitrary POVM and CP maps and classical communication, then all other allowed inputs (and their arbitrary convex combinations) satisfy $D_2(\rho^\mrin)=0$.

\noindent \textit{{Proof:}} Assume that some states in $\eL$ have $D_2(\rho^\mrin)\neq 0$. Introduce a CP map $\Phi(\rho)=G^+\big(G(\rho)\big)$. It is a unital map, because $G^+$ is unital \cite{karol}. According to Property 1 its application to $\rho_{00}:=|00\9\600|$ gives $\Phi(\rho_{00})=\rho_{00}$. Assume that Alice is the first party to perform a measurement on the inputs, and consider a state $\rho_{10}:=|10\9\610|$ (not necessarily an allowed input). Since $\Phi$ is unital,
\be
\Phi(\1-\rho_{00})=\Phi(\rho_{01}+\rho_{10}+\rho_{11})=\1-\rho_{00},
\ee
so the positivity of density matrices enforces $\60|\Phi(\rho_{10})|0\9=0$, and similarly for two other states in the preceding equation. As a result,  $\Phi(\rho_{10})$  has a disjoint support from $\rho_{00}$.

Separate the map $\Phi$ into Alice's first measurement $\{\Lambda_A^\mu\}$ and everything else. Evolution of any state $\rho^\mrin$ can be schematically written as $\rho^\mrin\mapsto\rho^\mu\mapsto\rho^\mout\mapsto\rho'$, with $\rho^\mout=U\rho^\mrin U^\dag$ for $\rho^\mrin\in\eL$, and $\rho'=\rho^\mrin$ for pure states in $\eL$. We write $\rho^\mu$ for $\rho_{|\Lambda^\mu_A}$ to simplify the notation. Since $\rho'=\Phi_\mu(\rho^\mu)$ for some CP map $\Phi_\mu$ by the lemma's assumption, and CP maps cannot improve  state distinguishability \cite{book, fuchs}, the post-measurement states $\rho^\mu_{00}$ and $\rho^\mu_{10}$ should have disjoint supports for any outcome $\mu$. Recall that in dealing with these two states Alice measures pure qubits while Bob's sides are identical. Hence Alice's measurement reduces to the projective measurement in some  basis (say $0'$,$1'$),
\be
\Lambda^a_A=\Pi_A^a=|a\9\6a|_A, \qquad a=0',1'.
\ee

Let Alice perform this  measurement on inputs  with  non-zero discord. For pure states  $\rho_A^\mrin$ the average post-measurement entropy becomes non-zero \cite{book, peres}. For mixed states with $D_2^A\neq 0$ Eq.~\eqref{ent_inc2} ensures that $S(\rho_\mrin^{\Pi_A})>S(\rho^\mrin)=S(\rho^\mout)$. However, projective measurements are repeatable, and a second measurement by Alice will certainly give the same result and induce no further change in the state. Hence, if the state $\rho^\mrin\in\eL$, then for any outcome $a$ the gate operates successfully, $G(\rho_{|\Pi^a_A}^\mrin)=G(\rho^\mrin)=U\rho^\mrin U^\dag.$  Since unitary maps preserve entropy and  and unital CP maps do not decrease it \cite{book, karol}, we reach a contradiction.

In case the first measurement is performed by Bob we consider the state $|01\9$ and use the discord $D_2^B$. \hfill $\blacksquare$

Now we consider what happens if the operation is performed on $d$-dimensional systems, and the set $\eL$ contains two non-orthogonal quantum states, $|\psi_i\9=|a_i\9|b_i\9$, $i=1,2$. Obviously as $|\psi'_i\9=U |\psi_i\9$,
\be
\6 a_1|a_2\9\6 b_1|b_2\9=\6 a'_1|a'_2\9 \6 b'_1|b'_2\9. \label{prodfix}
\ee
 This time we do not have to assume anything about the gate $G$ apart from its being implemented by LOCC. The following lemma explains our original example.

\textit{Lemma 2.} If the set $\eL$ contains two pure non-orthogonal states, and the unitary operation is such that
$D_2(\rho)\neq D_2(U\rho U^\dag)$, where $\rho=w\rho_{\psi_1}+(1-w)\rho_{\psi_2}$, for some $0<w<1$, then it cannot be implemented on $\eL$ by LOCC alone.

\noindent \textit{{Proof:}} Eq.~\eqref{prodfix} holds either through the constancy of the overlap on both sides individually, $|\6a_1|a_2\9|=|\6a'_1|a'_2\9|$, $|\6b_1|b_2\9|=|\6b'_1|b'_2\9|$, or by increasing one overlap and decreasing the other, as, for example, $|\6a_1|a_2\9>|\6a'_1|a'_2\9$, $|\6b_1|b_2\9|<|\6b'_1|b'_2\9|$. The latter possibility precludes LOCC gate execution, since the inequality $|\6a_1|a_2\9|>|\6a'_1|a'_2\9|$ entails that the distinguishability of two states improved as a result of some CP map, which is impossible \cite{book, fuchs}.

The product form of the final states makes it is possible  to find (non-unique) local unitary operations $U^i_A$, $U^i_B$ such that $|a'_i\9=U^i_A|a_i\9$, $|b'_i\9=U^i_B|b_i\9$. The norm conservation requires that when restricted to the linear spans of the states $|a_i\9$ and $|b_i\9$, respectively, these operators to satisfy $U^1_A=e^{i\alpha}U^2_A$ and $U^1_B=e^{i\beta}U^2_B$ for some phases $\alpha$ and $\beta$. As a result, on the states $\rho=w\rho_{\psi_1}+(1-w)\rho_{\psi_2}$ the gate is realized by a bi-local unitary operation,
\be
\rho'=U\rho U^\dag=U_A\otimes U_B\rho U_A^\dag\otimes U_B^\dag,
\ee
which implies \cite{disc-int,bt1} that $D_2(\rho)=D_2(\rho')$, contraindicating the assumption. Hence the LOCC implementation of $U$ is impossible. \hfill $\blacksquare$

\medskip

It is possible to draw several conclusions. First the absence of entanglement in both input and output does not automatically enable a remote implementation by LOCC.  Second, a discrepancy between local and global information content of  non-entangled states (which is captured by the discord $D_2$ in our setting and may have to be generalized in more sophisticated scenarios) requires entanglement for their processing. In the preceding cases  presented above we see that entanglement is required for any gate which changes the discord of the state. Recent results \cite{Concordant,EntDisc} suggest that a change in discord rather then entanglement is the required resource in computational speed-up. Our result shows that the two are intimately linked.

\medskip

We thank G. Brennen, A. Datta, R. Duan, F. Fanchini, K. Modi, J. Twamley, and  K. \.{Z}yczkowski for useful discussions and helpful comments.

\end{document}